\documentclass{article}
\usepackage{spconf,amsmath,graphicx,hyperref}
\usepackage{xcolor}
\usepackage{booktabs}
\usepackage{pifont} 
 \usepackage{amssymb}
\usepackage{subcaption}
 
\title{Dialogue-Aware Video-to-Music Generation Using Public Domain Film Collections}
\name{
Haven Kim$^{1}$ \quad
Zachary Novack$^{1}$ \quad
Juian McAuley$^{1}$ \quad
Hao-Wen Dong$^{2}$
}

\address{
$^{1}$University of California San Diego \\
$^{2}$University of Michigan
}
\begin{document}
%
\maketitle
\begin{abstract}
Video-to-music generation has drawn growing interest for its role in conveying the emotion of visual media, including film. Progress in the field, however, is hampered by a reproducibility gap: models are often trained on crawled corpora referenced through YouTube URLs that may be deleted, with the underlying data often difficult and time-consuming to retrieve. To address this, we introduce the \textbf{Open Screen Soundtrack Library version 2 (OSSL-v2)}, a self-hosted corpus of 34,343 video clips totaling 246.4 hours, sourced from public-domain films. Unlike crawled corpora, OSSL-v2 is reproducible (i.e., not subject to link rot) and copyright-conscious, yet still large enough to train functional video-to-music models. We then use this film-domain corpus to study dialogue as a conditioning signal for video-to-music generation, motivated by the close temporal coupling between film music and on-screen speech. Specifically, we augment existing models' video cross-attention with a time axis and modulate it frame-by-frame with the dialogue track. Evaluated on both public-domain and commercial films, our approach shows improvement over the state-of-the-art baselines. The dataset is available at \url{https://huggingface.co/datasets/McAuley-Lab/OSSL-v2}.

\end{abstract}
\section{Introduction}
\label{sec:intro}

Music is central to how video communicates: it conveys emotion, establishes mood, and shapes a viewer's interpretation of a scene~\cite{thao2021attendaffectnet, chua2022predicting}. Video-to-music generation systems, which automatically generate music that matches the emotional content of a video~\cite{boltz2004cognitive, thao2019multimodal, herget2021music, won2021emotion}, have therefore attracted growing interest and remain an important direction for research~\cite{gan2020foley,  su2024v2meow, tian2025vidmusesimplevideotomusicgeneration, lin2025vmas, zhang2025sonique, zuo2025gvmgen,  wang2025vision, cao2025ai, lin2026v2m, ji2026diff}. However, one problem stands out: the field lacks a common and durable training dataset that can serve as a benchmark.

The absence of such a dataset makes it difficult to measure progress in video-to-music generation, as most systems are trained on web-scraped corpora, often distributed as lists of YouTube or TikTok URLs~\cite{himv2017, tiktok2022, symmv2023, bgm909_2024, tian2025vidmusesimplevideotomusicgeneration}, for two reasons. First, the data is impermanent: videos are often removed or made private, meaning that models may be trained and evaluated on different, partially unavailable data, making faithful reproduction difficult. Second, the rate-limiting imposed by the host platforms substantially slows down obtaining the dataset. We observed this when attempting to retrieve a publicly released video-to-music benchmark~\cite{tian2025vidmusesimplevideotomusicgeneration} distributed as a list of YouTube IDs. We found that more than 10\% of its 300 benchmark videos were no longer accessible because they had been deleted or made private. Notably, this erosion occurred within only about two years of the dataset's initial release, and the inaccessible fraction can be expected to increase over time. Furthermore, the rate limiting from YouTube makes it time-consuming to re-crawl the full corpus of roughly 360k clips. Although a recent effort distributes self-hosted video-music pairs~\cite{kim2025video}, its scale remains limited: the dataset contains roughly 36.5 hours across 736 clips, making it difficult to use as a training and evaluation benchmark.

We build upon this self-hosted philosophy at scale by introducing the \textbf{Open Screen Soundtrack Library version 2 (OSSL-v2)}, a corpus of 34,343 video clips totaling approximately 246 hours, sourced from public-domain films. Because the dataset is self-hosted rather than referenced by URL, OSSL-v2 is not subject to deletion from third-party platforms and does not require time-consuming web scraping. At the same time, it is large enough to train competitive video-to-music generation models. OSSL-v2 therefore enables different research groups to train and evaluate on exactly the same data, making it suitable as a fixed benchmark for fair comparison.

  \begin{figure}[t]
    \centering
    \includegraphics[width=0.75\linewidth]{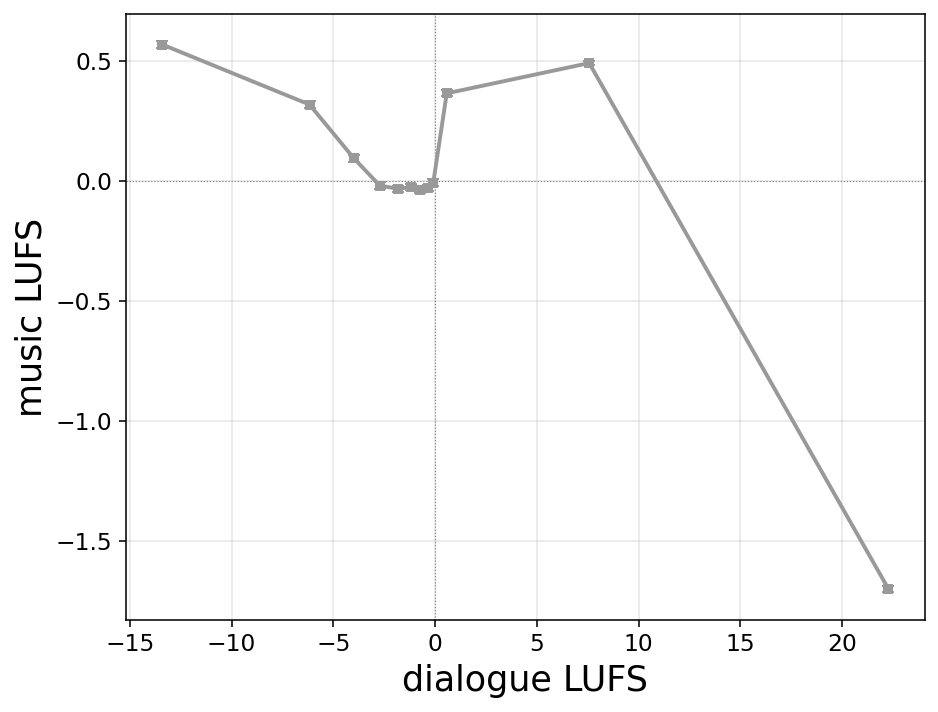}
    \caption{Binned ialogue vs.\ music loudness (LUFS) on OSSL-v2 ($r{=}-0.11$).}
    \label{fig:lufs}
  \end{figure}
  
Beyond data, we show that video-to-music generation may be missing a key input, particularly in the film domain: dialogue. In film, dialogue provides time-local cues about character interaction, narrative emphasis, and emotional density~\cite{banse1996acoustic}, and film music is often shaped around spoken lines ~\cite{torcoli2019background, addy2026improving} --- we observe that per-frame dialogue and music loudness (LUFS) are negatively correlated (Pearson $r{=}-0.11$; Fig.~\ref{fig:lufs}) in OSSL-v2 as well. These interactions are inherently local in time, and may be difficult to capture with global conditioning alone. 



To incorporate dialogue into video-to-music generation, we adapt existing open-source models~\cite{tian2025vidmusesimplevideotomusicgeneration, zuo2025gvmgen, ji2026diff} to use the speech track's acoustic envelope (e.g., loudness, energy) as a time-aligned conditioning signal during music generation. Because dialogue occurs at specific moments in a scene, the model must preserve when visual events and spoken lines occur relative to the generated audio. We therefore modify the video-conditioning pathway so that visual features retain temporal identity, the cross-attention memory can be aligned with the audio timeline, and dialogue can modulate the conditioning signal frame by frame. Trained and evaluated on OSSL-v2, this dialogue-aware formulation improves over state-of-the-art baselines on both public-domain and commercial film evaluation sets~\cite{kim2025video}, especially in terms of temporal fidelity, i.e., similarity to the ground truth.

\begin{table}[]
\caption{Comparison of paired video-music dataset. }
\centering
\resizebox{0.99\linewidth}{!}{
\begin{tabular}{@{}cccc@{}}
\toprule
Dataset & \begin{tabular}[c]{@{}c@{}}Self-\\ Hosted\end{tabular} & \begin{tabular}[c]{@{}c@{}}Video\\ Content\end{tabular} & \begin{tabular}[c]{@{}c@{}}Length\\ (Hours)\end{tabular} \\ \midrule
HIMV-200K~\cite{himv2017} & \color[HTML]{FE0000} \ding{55} & \begin{tabular}[c]{@{}c@{}}Music Video, User-Generated Video\end{tabular} & - \\
URMP~\cite{urmp2019} & \color[HTML]{FE0000} \ding{55} & Music Performance & 33.5 \\
TikTok~\cite{tiktok2022} & \color[HTML]{FE0000} \ding{55} & Dance Video & 1.5 \\
SymMV~\cite{symmv2023} & \color[HTML]{FE0000} \ding{55} & Music Video & 76.5 \\
MuVi-Sync~\cite{Kang_2024} & \color[HTML]{FE0000} \ding{55} & Music Video & - \\
BGM909~\cite{bgm909_2024} & \color[HTML]{FE0000} \ding{55} & Music Video & - \\
VidMuse~\cite{tian2025vidmusesimplevideotomusicgeneration} & \color[HTML]{FE0000} \ding{55} & Music Video, Advertisements, Trailer & 18k \\
OSSL~\cite{kim2025video} & \color[HTML]{009901} \ding{51} & Films & 36.5 \\
\textbf{OSSL-v2 (ours)} & \color[HTML]{009901}{\ding{51}} & \textbf{Films} & \textbf{246.4} \\
\bottomrule
\end{tabular}
}

\label{tab:comparison}
\end{table}

\section{Related Work}

In this section, we review prior work on video-to-music generation and their datasets, with an emphasis on music \emph{audio} generation rather than symbolic music. 

Early work in modern video-to-music generation systems uses large-scale web music-video corpora and autoregressive modeling over semantic acoustic tokens to generate music~\cite{su2024v2meow}. Subsequent work has advanced such systems along several dimensions. One direction improves visual representation learning and model architecture, including long- and short-term visual modeling~\cite{tian2025vidmusesimplevideotomusicgeneration}, hierarchical attention~\cite{zuo2025gvmgen}, dynamic motion or rhythmic modeling~\cite{liu2025extendingvisualdynamicsvideotomusic, ji2026diff}, a rhythm-aware adaptor~\cite{li2024vidmusicianvideotomusicgenerationsemanticrhythmic,li2024muvivideotomusicgenerationsemantic, tong2025videoechoedmusicsemantic}, parameter-efficient semantics~\cite{li2024vidmusicianvideotomusicgenerationsemanticrhythmic}, storyboard guidance~\cite{tong2025videoechoedmusicsemantic}, and semantic planning~\cite{lokegaonkar2026videorobinautoregressivediffusionplanning}. A second direction revisits training objectives and supervision by using beat-alignment objectives~\cite{lin2025vmas} or leveraging unpaired data, therefore relaxing paired supervision~\cite{zhang2025sonique, lin2026v2m}. Another direction introduces domain specialization, such as Chinese-style music generation~\cite{cao2025ai}. Finally, film-oriented approaches have been explored via composition-style transfer~\cite{qi2024harmonizingpixelsmelodiesmaestroguided}, a video-adapter to text-to-music generation model~\cite{kim2025video}, integration of different pipelines within a production-oriented framework~\cite{xie2025filmcomposerllmdrivenmusicproduction}. 

Beyond video-only conditioning, another line of work treats video-to-music generation as part of broader multi-modal music generation, enabling music creation from images, videos, text, or a combination of modalities~\cite{liu2024m2ugenmultimodalmusicunderstanding, liu2024mumullamamultimodalmusicunderstanding, wang2024multimodal}. 

On the other hand, a number of datasets have been developed for video-to-music generation tasks. These span various types of video content, including music videos~\cite{himv2017, symmv2023, li2024muvivideotomusicgenerationsemantic, bgm909_2024, tian2025vidmusesimplevideotomusicgeneration}, musical performance recordings~\cite{urmp2019}, and user-generated content~\cite{himv2017}. The most closely related work is the Open Screen Soundtrack Library~\cite{kim2025video}, which is a relatively small-scale dataset comprised of music-movie clip pairs sourced from public domain films. In short, most of the video-to-music datasets require a separate web-scraping procedure (i.e., not self-hosted), which makes the dataset prone to deletion and time-consuming to download, or its scale is not suitable for training purposes. Table~\ref{tab:comparison} compares our dataset against publicly available paired video--music datasets. Ours is among the few self-hosted video--music datasets, while also offering a relatively large scale.

\section{Open Screen Soundtrack Library version 2}

\noindent \textbf{Dataset Construction.} \indent Our video-music dataset, Open Screen Soundtrack Library Version 2 (OSSL-v2), is constructed from 1,886 public domain films downloaded from YouTube. The dataset is self-hosted, meaning that readers do not need to undergo a separate download process such as web scraping. Our dataset construction process consists of two main components: source separation and event detection. 

In the first step, in order to extract music from each movie clip's audio track, we apply an open source cinematic source separation model~\cite{solovyev2023benchmarks}, which offers a high-quality processing option that requires three times longer than the default option. We select the high-quality option because our objective is to create a music-movie clip dataset with the highest possible quality. In the second step, we employ an event detection model to estimate the probability distribution of event types in source-separated musical tracks. This step is essential because source-separated music, even when using a high-quality option, often contains non-musical events. To address this, we use an open-source automatic event detection model~\cite{kong2020panns}, from which we manually identify 157 out of 527 categories as musical events (e.g., ``trance music''). We define the music probability as the sum of probabilities for the 157 musical events, and the non-music probability as the sum of probabilities for the 370 non-musical events, to source-separated music. We extract segments where the music probability exceeds the non-music probability for at least 10 consecutive seconds. However, this fails to filter out cases where both musical and non-musical events are prominent (e.g., music probability of 0.8 and non-music probability of 0.7). Therefore, we apply an additional filter to exclude cases where the non-music probability exceeds a threshold, which we empirically set to 0.05.

This results in 34,343 movie clips paired with music, totaling 246.4 hours, with an average length of 28.6 seconds.

\begin{figure}[t]
    \centering

        \includegraphics[width=\linewidth]{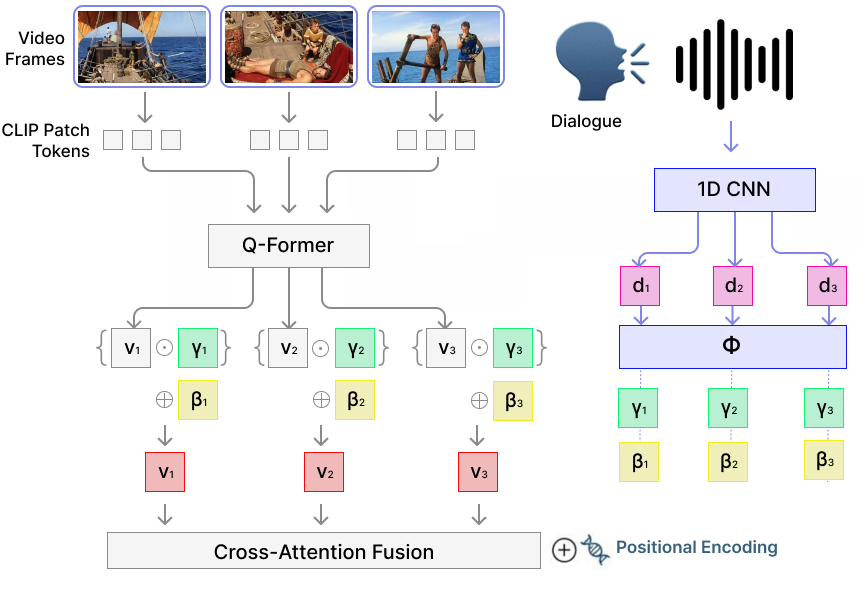}
        \caption{Overview of the proposed modification to GVMGen. Grayscale blocks correspond to the original GVMGen model, while colored blocks represent the added components.}
        \label{fig:modified}
\end{figure}

\section{Dialogue-Aware Video-to-Music Generation}
\label{sec:method}

In film, dialogue provides important cues about character interaction, narrative emphasis, and emotional intensity~\cite{banse1996acoustic}. These cues occur at specific moments in time, so a video-to-music model should be able to condition the generated score on when dialogue appears and how it aligns with the visual content, rather than relying only on global video-level information. To incoporate this information, we introduce a lightweight dialogue-conditioned module that can be inserted into pen-source video-to-music generation models. 


\noindent \textbf{Common Design} \indent Let the video conditioning of a backbone be a sequence of vectors $V=[v_1,\dots,v_N]$, where $v_i\in\mathbb{R}^{d}$ corresponds to video frame $i$. For each clip, we isolate the dialogue using the same source separation model used for dataset construction~\cite{solovyev2023benchmarks} and encode it with a small 1-D convolutional layer, yielding a per-frame low-level acoustic representation of dialogue (e.g., loudness, energy) $u_i \in \mathbb{R}^{c}$ temporally aligned to $V$. A two-layer MLP $\phi$ maps each dialogue representation to Feature-wise Linear Modulation (FiLM)~\cite{perez2018film} parameters $(\gamma_i,\beta_i)=\phi(u_i)$, which modulate the corresponding video-conditioning vector, 

\begin{equation}
  \tilde v_i \;=\; (1+\gamma_i)\odot v_i + \beta_i ,
  \label{eq:film}
\end{equation}

\noindent where $\odot$ denotes element-wise multiplication. The final layer of $\phi$ is zero-initialized, so that the module is initialized as an identity function at the start of training. How this condition is met differs across the backbones.

\noindent \textbf{VidMuse~\cite{tian2025vidmusesimplevideotomusicgeneration}} is a MusicGen~\cite{copet2023simple}-based autoregressive model that encodes video with a local branch, co-segmented with the target audio. Because it embeds each frame with CLIP~\cite{radford2021learning} along the time axis and adds a learned positional embedding, its cross-attention conditioning sequence is already temporarily order. We therefore apply Equation~\ref{eq:film} directly to the sequence. 

\noindent \textbf{GVMGen~\cite{zuo2025gvmgen}} is a MusicGen~\cite{copet2023simple}-based autoregressive Transformer that predicts discrete audio tokens conditioned on video through cross-attention. In the original model, each frame's CLIP~\cite{radford2021learning} patch tokens are compressed by a Q-Former~\cite{li2023blip} and then mean-pooled to a single vector per frame, and the resulting per-frame vectors are used as cross-attention conditioning without positional embeddings. Therefore, Applying Equation~\ref{eq:film} directly to this memory would collapse to a global effect, because the memory has no time axis. We therefore restore one with two further zero-initialized components. First, we add a learned per-frame positional embeddings to the Q-Former outputs before FiLM modulation (Equation~\ref{eq:film}), giving each frame a distinct positional identity. Second, we enable a fixed sinusoidal positional embedding on the cross-attention keys and values, allowing the audio-to-video attention to use this temporal identity. 

\noindent \textbf{Diff-V2M~\cite{ji2026diff}} is a Stable Audio Open~\cite{evans2025stable}-based latent diffusion model whose visual semantic features condition the generator through cross-attention on a temporal grid. We apply Equation~\ref{eq:film} to the per-frame semantic features, whith the FiLM module initialized as an identity function.

\section{Experiments}

The goal of our experiments are two-fold. First, we benchmark video-to-music generation tasks by comparing existing video-to-music generation models trained and evaluated on identical data, using OSSL-v2. Second, we assess the effectivenss of the proposed dialogue adapter when applied to these models.

\noindent \textbf{Compared Models.} \indent We comprehensively benchmark publicly available video-to-music generation models that learn from paired video-music data, namely VidMuse~\cite{tian2025vidmusesimplevideotomusicgeneration}, GVMGen~\cite{zuo2025gvmgen}, and Diff-V2M~\cite{ji2026diff}, together with the variants obtained by augmenting each model with the dialogue adapter described in Section~\ref{sec:method}. We exclude Sonique~\cite{zhang2025sonique} although it is an open-source video-to-music generation model, as it is designed to exploit unpaired data, making it incomparable under our protocol. For every model we adopt the exact hyperparameter settings recommended in the corresponding public repository, including the number of training steps, batch size, learning rate, etc.; where implementation details are missing from the released code, we follow the descriptions in the respective papers. One gap is the training segment length. Although all these models are originally trained on 30-second segments, we set the training length to 10 seconds, since the shortest clips in OSSL-v2 are 10 seconds long, and randomly crop a 10-second window from clips that exceed this length. At inference time, however, each model is required to generate a soundtrack matching the full duration of the input video clip.

\noindent \textbf{Dataset} \indent We partition OSSL-v2 into training and test splits with a 9:1 ratio, and all models are trained and evaluated on exactly the same splits. We report results on the test set as well as OES-Com~\cite{kim2025video}, a set of one hundred 30-second movie clips drawn from commercial films. The former reflects how well each model fits the training distribution, whereas the latter assesses generalization to out-of-distribution data. In addition, we evaluate on low-residual-music (LRM) subsets of both evaluation sets to assess whether any benefit from dialogue conditioning is merely an artifact of residual music leaking into the separated dialogue stem under imperfect source separation. Specifically, we run an audio event detection model~\cite{kong2020panns} on each clip's separated dialogue stem and retain only those clips whose summed probabilities over the 157 music event labels fall below 0.05 (i.e., the average probability is less than $\approx0.0003$). This yields 878 clips from the OSSL-v2 test set and 20 clips from OES-Com. If the dialogue adapter yields comparable improvements on these low-residual-music subsets and on the full evaluation sets, we can conclude that leftover music in the separated dialogue stem contributes little to the observed gains.

\noindent \textbf{Evaluation Metrics} \indent Following prior work~\cite{kim2025video}, we evaluate generation quality along three axes. We measure \emph{distributional fidelity}, i.e., how closely the distribution of generated music matches that of the reference music, using FAD~\cite{kilgour2018frfad} and Precision~\cite{naeem2020reliableqrdc}, as well as \emph{paired fidelity}, i.e., how semantically faithful each generated track is to its corresponding ground-truth music, using CLAP~\cite{wu2023large} similarity and the PaSST~\cite{koutini2021efficientpasst}-based KL divergence between the predicted audio-event distributions of the generated and reference tracks. Finally, we measure \emph{diversity} using Recall~\cite{naeem2020reliableqrdc}. Following prior work~\cite{kim2025video}, the reference-based metrics (FAD, Precision, and Recall) are computed against a reference music distribution drawn from 5,000 commercial film soundtracks that we privately crawled~\footnote{We release the precomputed reference embeddings’ mean and covariance, at \url{https://github.com/havenpersona/ossl-v2}.}

\begin{table*}[t]
\caption{Objective evaluation on the OSSL-v2 public-domain test set and the commercial OES-Com set, on the full evaluation sets (\emph{All}) and on the low-residual-music subsets (\emph{LRM}: clips whose separated dialogue stem has summed probability $<0.05$ over the 157 PANNs music-event labels, i.e.\ average probability $<0.0003$; 20/100 OES-Com clips and 878/3{,}332 OSSL-v2 clips). $\uparrow$ / $\downarrow$ indicate that higher / lower is better.}
\centering
\label{tab:results}
\resizebox{\textwidth}{!}{
\begin{tabular}{@{}lrrrrrrrrrrrrrrrrrrrr@{}}
\toprule
 & \multicolumn{5}{c}{\shortstack{OES-Com\\(All, n=100)}} & \multicolumn{5}{c}{\shortstack{OES-Com\\(LRM, n=20)}} & \multicolumn{5}{c}{\shortstack{OSSL-v2\\(All, n=3{,}332)}} & \multicolumn{5}{c}{\shortstack{OSSL-v2\\(LRM, n=878)}} \\
\cmidrule(lr){2-6} \cmidrule(lr){7-11} \cmidrule(lr){12-16} \cmidrule(lr){17-21}
 & Sim↑ & KL↓ & P↑ & FAD↓ & R↑ & Sim↑ & KL↓ & P↑ & FAD↓ & R↑ & Sim↑ & KL↓ & P↑ & FAD↓ & R↑ & Sim↑ & KL↓ & P↑ & FAD↓ & R↑ \\ \midrule
VidMuse~\cite{tian2025vidmusesimplevideotomusicgeneration} & 0.19 & 1.47 & 0.17 & 93.64 & 0.00 & 0.19 & 1.58 & 0.10 & 109.04 & 0.09 & 0.35 & 0.94 & 0.38 & 65.57 & 0.00 & 0.34 & 0.89 & 0.43 & 64.04 & 0.00 \\
(+Dialogue) & 0.23 & 0.85 & 0.04 & 97.70 & 0.03 & 0.22 & 0.89 & 0.10 & 113.24 & 0.00 & 0.36 & 0.82 & 0.32 & 66.12 & 0.00 & 0.35 & 0.80 & 0.45 & 66.89 & 0.00 \\
GVMGen~\cite{zuo2025gvmgen} & 0.23 & 0.97 & 0.53 & 73.40 & 0.17 & 0.23 & 0.94 & 0.55 & 89.81 & 0.03 & 0.43 & 0.72 & 0.56 & 49.05 & 0.22 & 0.43 & 0.72 & 0.56 & 48.98 & 0.25 \\
(+Dialogue) & 0.22 & 0.87 & 0.57 & 77.29 & 0.01 & 0.24 & 0.77 & 0.50 & 90.17 & 0.04 & 0.39 & 0.73 & 0.42 & 56.16 & 0.11 & 0.39 & 0.73 & 0.44 & 55.98 & 0.13 \\
Diff-V2M~\cite{ji2026diff} & 0.23 & 0.62 & 0.00 & 129.30 & 0.00 & 0.21 & 0.64 & 0.00 & 137.38 & 0.00 & 0.26 & 0.75 & 0.00 & 119.68 & 0.00 & 0.27 & 0.73 & 0.00 & 120.22 & 0.00 \\
(+Dialogue) & 0.25 & 0.63 & 0.00 & 130.72 & 0.00 & 0.23 & 0.68 & 0.00 & 137.45 & 0.00 & 0.27 & 0.68 & 0.00 & 118.87 & 0.00 & 0.28 & 0.64 & 0.00 & 119.41 & 0.00 \\
\bottomrule
\end{tabular}
}
\end{table*}
\section{Results}

Table~\ref{tab:results} reports the objective evaluation results for all compared models on both the public-domain test set and the commercial OES-Com set.

\noindent \textbf{Benchmarking video-to-music models.} \indent On the public-domain OSSL-v2 test set, GVMGen achieves the strongest overall performance, followed by VidMuse and Diff-V2M. Diff-V2M is a notable outlier in distributional fidelity: its Precision collapses to 0.00 and its FAD reaches a substantially higher value than the other models, indicating that its generated samples fall almost entirely outside the reference distribution. On the out-of-distribution OES-Com set, Diff-V2M attains the strongest paired fidelity, with the lowest KL and CLAP similarity on par with GVMGen. However, it again attains a Precision of 0.00, in contrast to the non-trivial Precision achieved by VidMuse (0.17) and GVMGen (0.53). We interpret this as evidence that Diff-V2M captures coarse semantic correspondences between video and music but fails to generate samples that conform to the ground-truth distribution. Taken together, these results establish GVMGen as the strongest model under our evaluation protocol.

\noindent \textbf{Effect of the dialogue adapter.} \indent We next examine the impact of augmenting each backbone with the proposed dialogue adapter (the ``+Dialogue'' rows in Table~\ref{tab:results}). The most consistent gains appear in paired fidelity. For VidMuse, the adapter sharply reduces KL divergence on both evaluation sets (OES-Com: 1.47$\to$0.85; OSSL-v2: 0.94$\to$0.82) while nudging CLAP similarity upward (OES-Com: 0.19$\to$0.23; OSSL-v2: 0.35$\to$0.36); Diff-V2M shows the same qualitative pattern, with CLAP rising on both sets (OES-Com: 0.23$\to$0.25; OSSL-v2: 0.26$\to$0.27) and KL improving on OSSL-v2 (0.75$\to$0.68). GVMGen is a partial exception: its paired fidelity improves on the out-of-distribution OES-Com set (KL: 0.97$\to$0.87) but degrades on the in-distribution OSSL-v2 set (CLAP: 0.43$\to$0.39; KL: 0.72$\to$0.73). Notably, the paired-fidelity gains are generally more pronounced on OES-Com than on OSSL-v2, suggesting that dialogue conditioning acts as a regularizer that generalizes beyond the training domain. In contrast, the adapter has no consistent effect on distributional fidelity: both Precision and FAD move in either direction across models and sets, indicating that dialogue conditioning does not systematically reshape the generated distribution itself.

The same behaviors persist on the low-residual-music (LRM) subsets, where the separated dialogue stem is almost free of leaked music. Across all three backbones, the direction and rough magnitude of every effect above are preserved on LRM---VidMuse's KL still drops sharply (OES-Com: 1.58$\to$0.89; OSSL-v2: 0.89$\to$0.80), GVMGen still improves on the OOD set (KL: 0.94$\to$0.77) while regressing on OSSL-v2, and its in-distribution paired fidelity still declines (CLAP: 0.43$\to$0.39). Because these behaviors remain when residual music is minimal, we attribute them to the dialogue signal itself rather than to music leaking into the dialogue stem.

\section{Conclusion}

We introduced OSSL-v2, a large-scale, self-hosted corpus of music-film clip pairs sourced from public-domain films. Because the dataset is free from link rot and does not require separate web scraping, our dataset is suitable as a durable benchmark for the field. 
Beyond the dataset, we showed that dialogue is an informative conditioning signal for music generation for film clips. By augmenting video-to-music generation models with frame-by-frame dialogue adapters along time axis, we obtain improvements over the corresponding baselines. We hope that OSSL-v2 lowers the barrier to reproducible research and fair comparison in video-to-music generation, and thereby supports the continued development of the filed, while our dialogue-aware experiments point to one promising direction for future work.

\section{Acknowledgement}
This work is partially supported by the NVIDIA Academic Grant Program under the project titled ``Teaser Generation for Long Documentaries and Educational Videos''.

\bibliographystyle{IEEEbib}
\bibliography{refs}

\end{document}